\pdfoutput=1
\documentclass[9pt,twocolumn,twoside]{pnas-new}
\usepackage{dsfont}
\usepackage{stackengine,scalerel,bbold}
\usepackage{ulem}
\usepackage{cuted}
\usepackage{mathrsfs}

\newcommand{\ee}{\operatorname{e}}

\newcommand{\RR}{\mathbb{R}}
\newcommand{\CC}{\mathbb{C}}

\newcommand\abs[1]{\left|#1\right|}
\newcommand\mLog[1]{-\log\left(\abs{#1}\right)}

\newcommand{\dif}{\operatorname{d}}

\templatetype{pnasresearcharticle} 

\title{Quantum tunneling and its absence in deep wells and strong magnetic fields}

\author[a,1]{Charles L. Fefferman}
\author[a,1]{Jacob Shapiro } 
\author[b,1]{Michael I. Weinstein}

\affil[a]{Department of Mathematics, Princeton University; Princeton, New Jersey USA}
\affil[b]{Department of Applied Physics and Applied Mathematics, and Department of Mathematics, Columbia University; New York, NY USA}

\leadauthor{Fefferman} 

\significancestatement{We establish surprising phenomena in magnetic systems, in strong contrast to the long-understood phenomena in non-magnetic systems. We construct quantum double well systems such that, in a strong magnetic field, the tunneling effect between wells is completely eliminated. Moreover, through appropriate deformation of the double well, the ground state can be made to transition from symmetric to anti-symmetric. This raises the question whether such phenomena could be observable in suitable experimental systems. Further, our results suggest an approach to achieving flat bands, which are of great interest in the study of strongly-correlated quantum systems. Some of our results rely on basic estimates on analytic functions, and could prove useful for other problems in mathematical physics.
}

\authorcontributions{All authors contributed equally to the research and writing of this article.}
\authordeclaration{No conflicts of interest.}
\correspondingauthor{\textsuperscript{2}To whom correspondence should be addressed. E-mail: author.two\@email.com}


\begin{abstract}
We present new results on quantum tunneling between deep potential wells, in the presence of a strong constant magnetic field. We construct a family of double well potentials containing examples for which the low-energy eigenvalue splitting vanishes, and hence quantum tunneling is eliminated. Further, by deforming  within this family, the magnetic ground state can be made to transition from  symmetric to anti-symmetric. However, for typical double wells in a certain regime, tunneling is not suppressed, and we provide a lower bound for the eigenvalue splitting. 
\end{abstract}

\dates{This manuscript was compiled on \today}
\doi{\url{www.pnas.org/cgi/doi/10.1073/pnas.XXXXXXXXXX}}

\begin{document}

\maketitle
\thispagestyle{firststyle}
\ifthenelse{\boolean{shortarticle}}{\ifthenelse{\boolean{singlecolumn}}{\abscontentformatted}{\abscontent}}{}


\dropcap{T}unneling is a phenomenon of central importance in quantum systems. An important paradigm is a quantum particle, such as an electron, in a double well potential. The ground state tunneling rate (tunneling probability) between weakly coupled wells is related to the magnitude of the energy splitting of the two lowest eigenvalues of the double well Hamiltonian: in a purely classical description the lowest two states of the system are degenerate. Physicists and chemists have long  known  \cite{Landau_Lifshitz_vol_3}, and mathematicians have proven \cite{Simon_1984_10.2307/2007072,Helffer_Sjostrand_1984}, that in the regime of deep potential wells (one way to realize the weakly-coupled regime), the energy splitting in non-magnetic quantum systems (a) is always non-zero and exponentially small in the depth of the wells, and (b) is asymptotically related to the \emph{hopping coefficient}, an overlap integral of ground state atomic orbitals. 

In this paper we consider two dimensional double well quantum systems in a constant perpendicular magnetic field. In \cite{FSW_22_doi:10.1137/21M1429412} we studied a Schr\"odinger double well model in which each single potential well is radially symmetric; see also \cite{helffer2023quantum,morin2023tunnelingeffectradialelectric}. The  constant magnetic field strength of order $\sim\lambda$ and the potential well depths $\sim\lambda^2$ are taken to be large.  We found (see Theorem 1) that (a) the eigenvalue splitting (and hence tunneling rate) is controlled by the hopping coefficient and (b) is always non-zero and exponentially small in $\lambda$.

In this article we announce new results on the general case, where the single well is no longer constrained to be radially symmetric.  We exhibit a family of double well potentials containing cases for which the low-energy eigenvalue splitting vanishes and quantum tunneling is eliminated. Further, by deforming  within this family, the magnetic ground state can be made to transition from  symmetric to anti-symmetric. For \emph{typical} double wells in a certain regime, we prove that tunneling is not suppressed, and we provide a lower bound for the eigenvalue splitting.
%
\subsection*{Mathematical setup} 
We consider a quantum mechanical particle in two space dimensions, with a constant magnetic field perpendicular to the plane. We shall study two related Hamiltonians: the one well Hamiltonian $h_\lambda$ and the two well Hamiltonian $H_\lambda$.
\subsubsection*{The magnetic one well Hamiltonian, $h_\lambda$}  
Let $x=(x_1,x_2)^\top$ and $x^\perp=(x_2,-x_1)^\top$. In the symmetric gauge for the magnetic vector potential, the one well magnetic Hamiltonian  is given by
     \[h_\lambda=\left(P+\frac12\lambda b X^\perp\right)^2 + \lambda^2 v(X) .\]
     Here, $Xf(x)=xf(x)$ and $P=-i\nabla$ denote the position operator and the momentum operator, respectively. We take the one well $v:\RR^2\to(-\infty,0]$  to be  smooth, nonzero and compactly supported within $B_a(0)$, the disc of radius $a>0$ in $\mathbb R^2$.
   The parameter $\lambda>0$, which sets the scale of both the depth of the potential and the magnetic field strength, is taken to be sufficiently large. The parameter $b>0$ is the relative (order $1$) strength of the magnetic field. For  fixed $b$, the effect of the magnetic field and that of a potential well with a non-degenerate minimum are of the same order of magnitude as $\lambda\to\infty$.
  
  Let $e_{\lambda}\sim -\lambda^2$ be the ground state eigenvalue of $h_\lambda$. Throughout this paper, we assume that (i)  $e_{\lambda}$ is simple, and that 
(ii)  $e_{\lambda}$ remains bounded away from  all other energies in the spectrum of $h_\lambda$ as $\lambda\to\infty$. 
 Assumption (ii) is valid if, for example, $v$ has an isolated non-degenerate minimum or if $v$ is very flat near its minimum, as for the square well. 

We denote the corresponding $L^2(\RR^2)$--normalized eigenfunction by $\varphi_\lambda(x)$; to simplify notation we will often write merely $\varphi(x)$. We emphasize that, in contrast to the non-magnetic case, it may not be possible to choose the ground state to be real-valued.

\subsubsection*{The magnetic double well Hamiltonian, $H_\lambda$} 
In some of the discussion below, we shall allow for $v(x)=v(x;\lambda)$, i.e., the one well potential depends on the large parameter $\lambda$. We replace the one well potential $v(x;\lambda)$ by two copies of $v$, one displaced by $d\in\RR^2$, the other by $-d$. 
The corresponding double well Hamiltonian is:
\begin{subequations}
    \begin{align} 
    H_\lambda &= \left(P+\frac12\lambda b X^\perp\right)^2  +  v^L(X;\lambda) + v^R(X;\lambda),\label{eq:Hlambda}\\
&\textrm{In the most basic case, we take\qquad\qquad\qquad\qquad\qquad\qquad\qquad\qquad\qquad }\nonumber\\    
    v^L(X;\lambda)&= \lambda^2 v(X+d)\quad  v^R(X;\lambda)=  \lambda^2 v(X-d).\label{eq:VLR}
    \end{align}
    \end{subequations}
Below, we shall also allow for more general $\lambda-$ dependence of   $v^L$ and $v^R$.

We center the two wells on the $x_1$--axis and impose that their supports do not overlap by taking:
\[ 
    d=\begin{bmatrix}d_1\\0\end{bmatrix},\quad d_1>a.\] 
If  $\lambda\gg1$, corresponding to the regime of deep wells and strong magnetic field, then under the constraints given above on $v$, $H_\lambda$ has two eigenvalues in a small neighborhood of the single well ground state energy, $e_{\lambda}$: 
     $E_{0,\lambda}\le E_{1,\lambda}$. The two-dimensional eigenspace corresponding to this pair of eigenvalues is approximately spanned by the (nearly orthonormal) symmetric and antisymmetric linear combinations of the ``atomic ground state orbitals'', $\varphi^L(x)$ and 
      $\varphi^R(x)$ arising separately from $v^L(x;\lambda)$ and $v^R(x;\lambda)$.
   The states $\varphi^L(x)$ and 
      $\varphi^R(x)$ are obtained via magnetic translation  of
      $\varphi(x)$ \cite{Zak_1964_PhysRev.134.A1602}:
      \[
    \varphi^L(x) = e^{i\frac{\lambda}{2} b d_1x_2} \varphi(x+d) \quad {\rm and}\quad 
    \varphi^R(x) = e^{-i\frac{\lambda}{2} b d_1x_2} \varphi(x-d).
      \]
    
     We introduce the \underline{magnetic eigenvalue splitting}: 
     \begin{equation}
   \Delta_\lambda := E_{1,\lambda}-E_{0,\lambda}
   \label{eq:SG}      
     \end{equation}
and the \underline{magnetic hopping coefficient}, an overlap integral of the single well ground state, $\varphi_\lambda(x)$,  magnetically translated to $x=-d$ and $x=+d$;  \cite{Zak_1964_PhysRev.134.A1602}:
\begin{subequations}
\label{eq:rho}
 \begin{align} 
 \rho_\lambda
 &\equiv\ \int \overline{\varphi^L(x;\lambda)}\  v^R(x;\lambda)\  \varphi^R(x;\lambda)\ dx. \label{eq:rho-def}\\
&\textrm{ In the most basic case [\ref{eq:VLR}], we have}\nonumber\\
\rho_\lambda &=  \int_{x\in\RR^2} \overline{\varphi(x+d)}\   \lambda^2 v(x-d) \ \varphi(x-d)\  e^{-i\lambda b d_1x_2}\ \mathrm{d}x\,,\label{eq:rho2}\end{align} 
 \end{subequations}
 Note that $\rho_\lambda$ is in general complex-valued. However, $\rho_\lambda$  is real if $ v^L(x;\lambda) + v^R(x;\lambda)$ is inversion symmetric (even).
 
For large $\lambda$, a formal analysis suggests that the emergent quantity,  $\rho_\lambda$, controls the eigenvalue splitting $\Delta_\lambda$, and hence tunneling between two wells. In non-magnetic systems ($b=0$) this has been proved. But as we shall see below, the magnetic case is far more subtle. 

The hopping coefficient also emerges in the strong binding regime for an infinite crystal Hamiltonian,  $H_{\rm crystal}$, whose potential is a superposition of atomic wells, centered on a discrete set $\mathbb G\subset\mathbb R^2$, in the sense that 
\begin{align}
	\left[ J 
    \left(H_{\rm crystal}-e_\lambda\mathds{1}\right) 
    J^*\right]_{nm} &\approx \rho_\lambda\  e^{i\frac{b\lambda}{2} n\wedge m} \delta_{\|m-n\|,d_0},\quad n,m\in\mathbb G.
	\label{eq:crystal-limit}
 \end{align} 
 Here, $d_0$,  the minimal distance within $\mathbb{G}$, is strictly positive and the minimal next-nearest-neighbor distance is uniformly bounded away from $d_0$. The operator $J$  relates functions on the continuum with vectors whose amplitudes are indexed by $\mathbb{G}$; see \cite{ShapWein22}.

 \subsection*{Previous work}
 Upper bounds for the eigenvalue splitting and the magnitude of the  
hopping coefficient are straightforward.
 Concerning previous work on lower bounds:

\begin{itemize}
    \item  \underline{Non-magnetic case ($b=0$)}: Under the condition that $v$ has a single non-degenerate minimum, it was proved in \cite{Simon_1984_10.2307/2007072,Helffer_Sjostrand_1984} that $\Delta_\lambda$ satisfies an exponential lower bound; 
    the relationship between $\Delta_\lambda$ and $\rho_\lambda$, and an asymptotic expansion of $\rho_\lambda$ was given in \cite{Helffer_Sjostrand_1984}.  In \cite{FLW17_doi:10.1002/cpa.21735}, a lower bound on $\rho_\lambda$ was proved with no assumption on the minimum of $v$, but assuming $v$ is of compact support.
\item \underline{Perturbative magnetic case ($b\sim\frac{1}{\lambda}$)}: Lower bounds on $|\rho_\lambda|$ have been obtained under the assumption of a weak magnetic field,  where $b \sim \frac{1}{\lambda}$ in \cite{Helffer_Sjostrand_1987_magnetic_ASNSP_1987_4_14_4_625_0}. This regime is perturbative, relative to the non-magnetic case, since  the magnetic field strength is of order $1$ and the potential wells are deep, of order $\lambda^2$.
\end{itemize}

The first result providing lower bounds for deep wells and strong magnetic fields was proved in \cite{FSW_22_doi:10.1137/21M1429412}: \medskip

\noindent {\bf Theorem 1:} 
{\it Magnetic hopping and magnetic splitting for radially symmetric atomic potentials.} Assume that $v$ is smooth and radial, such that $\varphi_\lambda$ is also radial. Consider the hopping, $\rho_\lambda$ and splitting $\Delta_\lambda$ associated with the magnetic double well Hamiltonian [\ref{eq:Hlambda}-\ref{eq:VLR}]. There exists $d_\star>0$, $c>0$ and $\lambda_\star>0$ such that for all $d_1>d_\star$: 
\begin{align*}
&-\rho_\lambda \ge\ \mathrm{e}^{-c\lambda\|d\|^2}\ \textrm{for  $\lambda\geq\lambda_\star$},\quad {\rm and}\\  
{\ }\\
 & \lim_{\lambda\to\infty} \quad \frac{\Delta_\lambda}{2|\rho_\lambda|} \to 1 .
 \end{align*}

  The articles \cite{helffer2023quantum,morin2023tunnelingeffectradialelectric} sharpen Theorem 1 by providing leading order asymptotics of $\rho_\lambda$ for $\lambda\to\infty$. 

The radial symmetry assumption on $v$ in Theorem 1 is very restrictive; we used it to rewrite the expression for $\rho_\lambda$ as a real integral with non-oscillatory phase, to which we applied Laplace asymptotics for $\lambda$ large.

\subsection*{The Non-Radial Case}
For double well systems, where the single well is not radially symmetric,  the situation is surprisingly different. \medskip

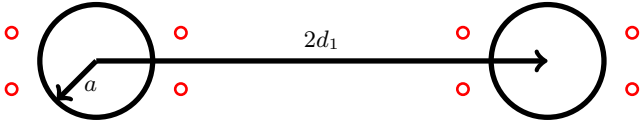
\begin{figure}
	    \centering
	    \begin{tikzpicture}[x=1cm,y=1cm,scale=0.75]
\draw [line width=2pt] (0,0) circle (1cm);
\draw [line width=1pt,red] (1.5,0.5) circle (0.1cm);
\draw [line width=1pt,red] (1.5,-0.5) circle (0.1cm);
\draw [line width=1pt,red] (-1.5,0.5) circle (0.1cm);
\draw [line width=1pt,red] (-1.5,-0.5) circle (0.1cm);
\draw [line width=2pt] (8,0) circle (1cm);
\draw [line width=1pt,red] (9.5,0.5) circle (0.1cm);
\draw [line width=1pt,red] (9.5,-0.5) circle (0.1cm);
\draw [line width=1pt,red] (6.5,0.5) circle (0.1cm);
\draw [line width=1pt,red] (6.5,-0.5) circle (0.1cm);
\draw [->,line width=2pt] (0,0) -- (8,0);
\draw [->,line width=2pt] (0,0) -- (-0.7,-0.7);
\node[below] at (-0.1,-0.2) {$a$};
\node[below] at (4,0.7) {$2d_1$};

\end{tikzpicture}
	    \caption{Inversion-symmetric potential of the type used in the proof of Theorem 2. Large circles are the supports, $B_0^L$ and $B_0^R$,   of $v_0^L(x)$, centered at $x^L_0=(-d_1,0)$, and $v_0^R(x)$, centered at $x^R_0=(+d_1,0)$.  
        For $\alpha=L,R$, we add four potentials (``sophons''), which are inversion symmetrically placed relative to $x_0^\alpha$ .  The sophons are supported on the small red discs, $B_\mu^\alpha$, which are centered at $x_\mu^\alpha$; $\mu=1,2,3,4$,
        the vertical components of which determine the phase $\theta$ in [\ref{eq:rho-asymp}]; the amplitudes, 
         support radii and centers of the sophon potentials are specially tuned   $\lambda-$ dependent functions.  }
	    \label{fig:setup for double well}
	\end{figure}

\noindent {\bf Theorem 2:} {\it (Magnetic double wells with zero tunneling)} Fix a radial one well potential $v_0(x)$ satisfying the hypotheses of Theorem 1. Then there is a $\lambda$-dependent family $\mathcal{F}_\lambda$ of (non-radial) one well potentials, such that the following hold for all large enough $\lambda$: 
\begin{enumerate}
    \item \emph{Potentials in $\mathcal{F}_\lambda$ are exponentially close to $\lambda^2v_0$}: Let $v(\cdot;\lambda)\in\mathcal{F}_\lambda$. Then $|\lambda^2v_0(x)-v(x;\lambda)|<\exp\left(-C\lambda\right)$ for all $x\in\mathbb{R}^2$. Moreover, $v(x;\lambda)\neq \lambda^2 v_0(x)$ only for $x$ in a set of area less than $\exp(-C \lambda)$. Here, $C>0$ is determined by $v_0$.
    \item \emph{Inversion symmetry}: 
    The two well potential \[v^L(x;\lambda)+v^R(x;\lambda) = v(x+d;\lambda)+v(x-d;\lambda)\] is inversion-symmetric for all $v\in\mathcal{F}_\lambda$.
    \item \emph{Absence of quantum tunneling for $H_\lambda$}: The family $\mathcal{F}_\lambda$ contains a potential for which the eigenvalue splitting vanishes, i.e., $\Delta_\lambda=0$. Hence, no tunneling occurs for such double wells. Moreover, $\mathcal{F}_\lambda$ also contains a potential for which the hopping coefficient $\rho_\lambda=0$.
    \item \emph{Odd ground state}: By deforming within the family $\mathcal{F}_\lambda$, the two well ground state eigenfunction of $H_\lambda$, $\psi$,  can be made to transition between symmetric ($\psi(x)=\psi(-x)$) and antisymmetric ($\psi(x)=-\psi(-x)$). 
\end{enumerate}
Figure \ref{fig:setup for double well} is a schematic showing the support of a potential in $\mathcal{F}_\lambda$. Such a potential is a perturbation of the double well potential of Theorem 1, which is comprised of two radially symmetric single wells. A discussion of the proof, with the aid of Figure \ref{fig:setup for double well}, is given below. \medskip 

We do not have an example of a non-inversion symmetric potential for which $\Delta_\lambda=0$.
However, we have the following result.
\medskip

\noindent {\bf Theorem 3:} 
There exist non-inversion symmetric magnetic double well potentials for which $\rho_\lambda=0$.
\medskip


The potentials $v(\cdot;\lambda)$ realizing zero tunneling or zero hopping  
coefficient in Theorems 2 and 3 do not arise by simply scaling a  
$\lambda$-independent potential.
We conjecture that a potential $v_0$, nonradial and independent of $\lambda$,
can be chosen so that the double well Hamiltonian arising from
$\lambda^2v_0$ gives rise to zero eigenvalue splitting for an infinite
sequence of $\lambda$ values tending to infinity, and similarly for the hopping
coefficient.
\medskip

We next claim that, in a precise sense, the
counterexamples of Theorems 2 and 3 are exceptional in a regime where $b$ (the magnetic strength parameter in [\ref{eq:Hlambda}]) is small but fixed as $\lambda\to\infty$. 
And further, in that regime, for typical large $\lambda$, there is a quantitative lower bound on both $|\rho_\lambda|$ and $\Delta_\lambda$. Recall that a set $S\subseteq(0,\infty)$ has  {\it zero density} if $\mu(S\cap[0,M])/M\to0$ as $M\to\infty$, where $\mu$ denotes Lebesgue measure.\medskip

     {\bf Theorem 4 :}  Fix $\gamma\in(0,1)$. Suppose the one well potential $v(x)$ is $C^3$ and has a unique minimum at the origin, with its Hessian there obeying $v''(0) = \mathds{1}_{2\times2}$. 
     Consider the magnetic double well operator [\ref{eq:Hlambda}-\ref{eq:VLR}]. Suppose $|b|<c_\gamma$ for some small constant $c_\gamma$. Then for all $\lambda$ outside a set $\mathcal{Z}(\gamma,v)\subseteq(0,\infty)$ of density zero, we have \begin{equation}\label{eq:lower bound on rho and Delta}
     |\rho_\lambda|>\exp(-\lambda^{\frac{1}{\gamma}})\qquad\textrm{and}\qquad  \Delta_\lambda > \exp(-\lambda^{\frac{1}{\gamma}})\,.\end{equation}

     Note that here, although $b$ is assumed small, it remains independent of $\lambda$, which may be taken arbitrarily large. Recall that our two well potential is $\lambda^2(v(x+d)+v(x-d))$ and the magnetic field strength is $b\lambda$. 
Finally, we expect the condition $v''(0) = \mathds{1}_{2\times2}$ can be relaxed.

\subsection*{Flat band materials}
The ideas of  
Theorems 2 and 3 can be applied to the design of periodic structures, which have nearly flat band spectrum in a neighborhood of the one well ground state energy. We explain this in a section below.

 \subsection*{Ideas underlying the proofs} 

 Let us perturb a radial one-well potential $v_0$ by adding a small correction $\delta v$. Then the hopping coefficient $\rho$ will change by an amount $\delta\rho$. Due to the strong magnetic field $(\lambda\gg1)$, $\delta\rho$ is given by an oscillatory integral. However, if the support of $\delta v$ has sufficiently small diameter (depending on $\lambda$), then the expected cancellation due to the rapidly oscillating exponential in the relevant integrand will not occur. 
Moreover, due to the rapid (Gaussian) decay
of the ground state for $v_0$,  the effect of $\delta v$ on $\delta\rho$ is enhanced if the
support of $\delta v$ is favorably placed. For instance, in Figure \ref{fig:setup for double well}, two of
the small discs on the left are closer than the large disc on the left
to everything on the right. 
 Consequently, although $\delta v$ is small compared to $v_0$, it may happen that $\delta\rho$ is large compared to $\rho$. We find that we can arrange $\delta\rho$ to have greater, smaller, or the same order of magnitude as $\rho$, as we please. This intuition plays a role in the proofs of Theorems 2 and 3. Here, we focus on the proof of Theorem 2.
 

\section*{Sketch of proof of Theorem 2}
 We refer to Figure \ref{fig:setup for double well}. We denote the large disc (a  ``planet'') on the left by $B_0^L$, centered at $x_0^L=(-d_1,0)$ and the four small discs on the left by $B_\mu^L$ with $\mu=1,\dots,4$; ( ``sophons'' \cite{liu2014three}); similarly for the right with $L\to R$ and $B_0^R$ centered at $x_0^R=(d_1,0)$. The centers of these discs are denoted as $x_\mu^\alpha$ for $\alpha=L,R;\quad\mu=0,\dots,4$. The centers, $x_0^L$ and $x_0^R$ are independent of $\lambda$. 
 The centers of the small discs, $x_\mu^\alpha$ for $\alpha=L,R;\quad\mu=1,\dots,4$, and the radii of $B_\mu^\alpha, \mu=1,2,3,4$ and $\alpha=L,R$ 
are chosen to depend on $\lambda$. The radii of these discs are chosen to  shrink exponentially fast as $\lambda\to\infty$.

Our  ``two-well'' potential is $v^L(x;\lambda)+v^R(x;\lambda)$, where 
\[ v^\alpha(x;\lambda) =  \lambda^2 v_0^\alpha(x) + \sum_{\mu=1}^4 v_\mu^\alpha(x;\lambda),\quad \alpha=L,R.\] Here, 
 (a) $v_0^\alpha(x)$ is independent of $\lambda$, and supported on  $B_0^\alpha$ (the planet), and (b) for $\mu=1,\dots,4$,  $v_\mu^\alpha(\cdot ;\lambda)$ is supported on $B_\mu^\alpha$ (a sophon), and $\|v_\mu^\alpha\|_\infty \sim \tau$  with $\tau=\tau(\lambda)$ a carefully chosen  constant which is exponentially small in $\lambda$.

 For $\alpha=L, R$, we write $\varphi^\alpha$ to denote the ground state of the one-well magnetic Hamiltonian $h^\alpha = \left(P+\frac12\lambda b X^\perp\right)^2+v^\alpha(X;\lambda)$. 
%
We study the magnetic hopping coefficient $\rho_\lambda$, defined in [\ref{eq:rho-def}]. Due to inversion symmetry, $\rho_\lambda$ is real. Since $v^R$ is supported on the five disjoint discs $B^R_\nu$, we have $$ \rho = \sum_{\nu=0}^4 \int_{B_\nu^R} \overline{\varphi^L} v^R \varphi^R\,. $$ 
Carefully justified 
perturbation theory lets us write 
\[  \varphi^L =  \varphi_0^L + \sum_{\mu=1}^4 \varphi_\mu^L \quad 
\]   modulo a negligible error. Here,  $\varphi_0^L$ denotes the magnetic ground state arising from $\lambda^2 v_0^L(x)$ while, for $\mu=1,2,3,4$, the function $\varphi^L_\mu$ denotes the correction to $\varphi^L_0$ arising from the added potential $v^L_\mu(x;\lambda)$ in first order perturbation theory.  Hence, modulo  a negligible error, we find 
\begin{align} \rho_\lambda &=\sum_{\mu=0}^4\left[\lambda^2\int_{B_0^R} \overline{\varphi^L_\mu}\ v^R_0\ \varphi^R+\sum_{\nu=1}^4 \int_{B_\nu^R} \overline{\varphi^L_\mu}\ v^R_\nu\ \varphi^R\right] \\
&=: \sum_{\mu,\nu=0}^4 \mathrm{T}(\mu,\nu)\,. \nonumber
\label{eq:rho-expand}
\end{align}
Each term, $T(\mu,\nu)$, may be thought of as a contribution to the tunneling coming from an interaction between planets and sophons on the left, $L$, and right, $R$. In particular, $T(0,0)$ corresponds to an interaction between planets; it is the hopping coefficient for the unperturbed
radial one well potential, $\lambda^2 v_0(x)$, to which Theorem 1 applies. The terms $T(\mu,\nu)$, with $\mu+\nu\ge1$, are corrections due to the introduction of sophons.

Now eigenstates $\varphi$ of the one well Hamiltonian satisfy $\left((P+\frac12\lambda b X^\perp)^2-e_\lambda\right)\varphi=0$ outside of the support of $v$, and hence  have Gaussian decay. Using this, we find that as $\lambda\to\infty$ 
\begin{subequations}
\label{eq:Tmunu}
\begin{align}
        |T(0,0)| &\sim  \exp\left(-\frac{\lambda}{4}\|x^L_0-x^R_0\|^2(1+o(1))\right)\\
        |\mathrm{T}(\mu,\nu)| &\sim \tau\exp\left(-\frac{\lambda}{4}\|x^L_\mu-x^R_\nu\|^2(1+o(1))\right),\\
      &  \textrm{where $\mu=0$ and $\nu\in\{1,2,3,4\}$} \nonumber\\
      & \textrm{or $\nu=0$ and $\mu\in\{1,2,3,4\}$} \nonumber\\
        |\mathrm{T}(\mu,\nu)| &\sim \tau^2\exp\left(-\frac{\lambda}{4}\|x^L_\mu-x^R_\nu\|^2(1+o(1))\right),\\\nonumber
        &\textrm{where $\mu, \nu \in\{1,2,3,4\}$}.
\end{align}
\end{subequations}

%
%

Here we have used the fact that the diameter of the supports of 
$v_\mu^L(x)$ and $v_\mu^R(x)$ shrinks to zero for $\mu=1,2,3,4$ at an exponential rate as $\lambda\to\infty$. 
Due to the effects discussed above in the section: {\it Ideas underlying the proofs}, we may choose $\tau=\tau(\lambda)$ to be exponentially small in $\lambda$, and such that
$T(0,\mu)$ and $T(\mu,0)$\ ($\mu=1,2,3,4$), which are formally of order $\tau$, dominate  $T(0,0)$, the unperturbed radial hopping coefficient. 
Further, one can derive from the expression for $\rho_\lambda$ in [\ref{eq:rho2}] that the vertical displacements of the sophons (see Figure \ref{fig:setup for double well}) introduce complex phases into $T(\mu,\nu)$ for $(\mu, \nu) \ne (0,0)$. The relative phase, $\theta$, of the dominant terms depends sensitively on the dimensions of the rectangle formed by $x_1^R,\dots,x_4^R$. 
   Accordingly, [\ref{eq:Tmunu}] implies that    
\begin{equation}
\rho_\lambda = \tau\ \exp\Big(-\frac{\lambda}{4}D^2\left(\ 1+o(1)\ \right) \Big) \big[\cos(\theta)+o(1)\big],\quad \textrm{for $\lambda\gg1$,} 
    \label{eq:rho-asymp}
\end{equation}
where $D:=\min_{1\le\mu\le4}\|x_\mu^L-x_0^R\|$. Here, the cosine appears in [\ref{eq:rho-asymp}] due to inversion symmetry, which dictates that $\rho$ is real.
From [\ref{eq:rho-asymp}]  we see that by slightly varying the placement of the sophons we can arrange for $\rho_\lambda$ to be either positive or negative. 

Let us now consider the spectral gap $\Delta_\lambda$. Since the two-well potential $v^L+v^R$ is even, the magnetic Hamiltonian $H_\lambda = \left(P+\frac12\lambda b X^\perp\right)^2+v^L+v^R$ commutes with the parity operator $f(x)\mapsto (\mathcal{P}f)(x) = f(-x)$, and hence preserves the subspaces $L^2_\mathrm{even}$, $L^2_\mathrm{odd}$ consisting of even wavefunctions $\psi(x)=\psi(-x)$ and odd wave functions $\psi(x)=-\psi(-x)$, respectively. Let $E_{\mathrm{even},\lambda}$ and $E_{\mathrm{odd},\lambda}$ denote the ground state eigenvalues of $H_\lambda$ restricted to $L^2_\mathrm{even}$ and $L^2_\mathrm{odd}$, respectively.
Then, for the lowest two eigenvalues of $H_\lambda$, we have  $\{E_{0,\lambda}, E_{1,\lambda}\} =
\{E_{\mathrm{even},\lambda} , E_{\mathrm{odd},\lambda}\}$. Hence, by [\ref{eq:SG}], the spectral gap of $H$ is then 
$$\Delta_\lambda = |E_{\mathrm{even},\lambda}-E_{\mathrm{odd},\lambda}|\,.$$ 
Moreover, modulo a small error, $E_{\mathrm{even},\lambda}-E_{\mathrm{odd},\lambda} = 2\rho_\lambda$. By deforming our potential (by moving the discs $B_\mu^\alpha$, $\mu=1,2,3,4$, $\alpha=R,L$) we can, by [\ref{eq:rho-asymp}] and the ensuing discussion, continuously vary $E_{\mathrm{even},\lambda}-E_{\mathrm{odd},\lambda}$ from a positive value to a negative value. Consequently, \begin{itemize}
    \item At some point during the deformation, we have $E_{\mathrm{even},\lambda}=E_{\mathrm{odd},\lambda}$, and hence $\Delta_\lambda=0$.
    \item We may take $E_{\mathrm{even},\lambda}-E_{\mathrm{odd},\lambda}$ to be positive, i.e., the ground state of the magnetic double well may be odd rather than even.
\end{itemize}
The class $\mathcal{F}_\lambda$ in Theorem 2 consists of all one-well potentials arising as above, for all possible placements of the four ``sophons'', $B^L_\mu$ ($\mu=1,2,3,4$) in a rectangle about the ``planet'' $B_0^L$, as in Figure \ref{fig:setup for double well}. This completes our rough sketch of the proof of Theorem 2.

\medskip

\subsection*{Sketch of proof of Theorem 4:} We study the ground state $\varphi_\lambda$ of the one well Hamiltonian $h_\lambda$, and the orthogonal projection $\Pi_\lambda$ from $L^2(\mathbb{R}^2)$ to the two lowest eigenstates of the two well Hamiltonian $H_\lambda$. We show that the $L^2$-valued function $\lambda\mapsto\varphi_\lambda$ and the operator-valued function $\lambda\mapsto\Pi_\lambda$ can be continued to analytic functions on the region 
\begin{equation}
\Gamma_\gamma := \{\Re{\lambda} > C_1 ,\,|\arg(\lambda)|<\gamma\ \pi/2\},\label{eq:Gamma-gamma}
\end{equation}
where $C_1<\infty$ is a large fixed constant and $0<\gamma<1$, provided $|b|<c_\gamma$, a sufficiently small and positive constant independent of $\lambda$. From this we conclude that $\lambda\mapsto\rho_\lambda$ and $\lambda\mapsto(\Delta_\lambda)^2$ also continue to analytic functions on $\Gamma_\gamma$. 
We first prove Theorem 4 assuming the analyticity of $\lambda\mapsto\rho_\lambda$ and $\lambda\mapsto(\Delta_\lambda)^2$. Then, in a section below we outline the proof of analyticity.

The strategy for deducing Theorem 4 from analyticity is to combine (i) upper bounds on the above analytic functions in $\Gamma_\gamma$ with,  (ii) a lower bound on these functions at a single point in $\Gamma_\gamma$ and then apply the complex analysis theorem below to prove an averaged lower bound on on $|\rho_\lambda|$ and $\Delta_\lambda$. This average lower bound implies that the lower bounds in [\ref{eq:lower bound on rho and Delta}] hold except for a set of density zero.

The upper bounds (i) follow from our analysis of $\lambda\mapsto\varphi_\lambda$ and $\lambda\mapsto\Pi_\lambda$ for complex $\lambda\in\Gamma_\gamma$, promised above. To explain our derivation of the lower bound (ii), we now write $\Delta_\lambda\equiv\Delta(\lambda,b)$ to denote the spectral gap arising from the Hamiltonian $H_\lambda$ for a given $\lambda$ and $b$. First set $b=0$.
 Then, $\Delta(\lambda,0)$ is the spectral gap for the non-magnetic Hamiltonian. By classical results for the non-magnetic Hamiltonian (see, for example, \cite{Helffer_Sjostrand_1984,Simon_1984_10.2307/2007072,FLW17_doi:10.1002/cpa.21735}), for fixed real $\lambda_0> C_1$ (see [\ref{eq:Gamma-gamma}]), we have $\Delta(\lambda_0,0)>\delta_1$ for some positive constant $\delta_1$. Consequently, for $|b|$ small enough -- say, $|b|<b_0(\lambda_0)$ -- we have $\Delta(\lambda_0,b)>\frac12 \delta_1$. This is the required lower bound (ii) for $|\Delta(\lambda,b)|^2$ for a single value of $\lambda$ and small $b$. 
 
 To deduce Theorem 4 from the above bounds (i) and (ii), we apply the following result to $z\in\Lambda_\gamma\mapsto \rho_z$ and 
 $z\in\Lambda_\gamma\mapsto \Delta(z,b)^2$, for $|b|<b_0(\lambda_0)$:  


\paragraph{A theorem in complex analysis} Let  $\Gamma_\gamma$ be the region defined in [\ref{eq:Gamma-gamma}], for $\gamma\in(0,1]$. 
Assume that $F:\Gamma_\gamma\to\CC$ is analytic and such that there exists $\beta>0$ for which \begin{align}\abs{F(\lambda_0)}\geq\ee^{-\beta},\end{align}
for  some real $\lambda_0>C_1$.
Further assume that there exists a non-vanishing analytic function $U:\Gamma_\gamma\to\CC$ such that for all $z\in\Gamma_\gamma$, 
\begin{align}
            \abs{F(z)} \leq \abs{U(z)}.
        \end{align}
        There exists a constant $C$ (depending on $\beta,\gamma$, $\lambda_0$ and $U$) such that for all R  
sufficiently large,
        \begin{align}
            \frac{1}{R}\int_{t=R}^{2 R}\mLog{F(t)}\dif{t}&\leq C R^{1/\gamma}+\\
            &\qquad+\frac{1}{R}\int_{t=R}^{2 R}\mLog{U(t)}\dif{t}\,.
        \end{align}
The above theorem is an application of the standard Blaschke factorization of analytic functions on the disc, see Theorem 15.21 of \cite{rudin1987real}.

\noindent {\it The proof of analytic continuation:}
 Making the above analytic continuations is not straightforward, since for typical $\mu\in\mathbb{C}$ the resolvents $(h_\lambda-\mu\mathds{1})^{-1}$ and $(H_\lambda-\mu\mathds{1})^{-1}$ do \emph{not} admit analytic continuations as functions of $\lambda$ taking values in the space of bounded operators on $L^2$.
However, when applied to functions supported on $B_a(0)$, these resolvents do admit analytic continuation to $\lambda\in\Gamma_\gamma$. Once this is proved, it is then not hard to control the above ground state $\varphi_\lambda$ and projection $\Pi_\lambda$ as functions of complex $\lambda\in\Gamma_\gamma$, where $0<\gamma<1$.

To establish the analyticity of $\lambda\mapsto (h_\lambda-\mu\mathds{1})^{-1}\chi$, where ${\rm supp}(\chi)\subset B_a(0)$, we fix $\Lambda\gg1.$
 For $\lambda\in\Gamma_\gamma$, with $\frac12\Lambda\le|\lambda|\le2\Lambda$, we localize $\chi$ into the regions \[ U_{\rm inner}= \big\{|x|\le \Lambda^{-\frac12+\varepsilon}\big\}\ {\rm and}\ 
 U_k= \big\{2^{-(k+1)}\le|x|\le 2^{-(k-1)}\big\},\] 
for $|\Lambda|^{-\frac12+\varepsilon}\le2^{-k}\le a$. In the region $U_{\rm inner}$, $h_\lambda$ is closely approximated by a magnetic harmonic oscillator Hamiltonian. 
Consequently, for ${\rm supp}(\chi)\subset U_{\rm inner}$, we can control 
$(h_\lambda-\mu\mathds{1})^{-1}\chi$ in terms of the resolvent of the magnetic harmonic oscillator (MHO):
\[ H^{\mathrm{MHO}}_\lambda = (-\mathrm{i}\nabla + \frac12 \lambda b X^\perp)^2+\frac12\lambda^2\omega_0^2X^2.\]
The resolvent kernel of $H^{\mathrm{MHO}}_\lambda$ can be expressed via the Laplace transform of the (magnetic) Mehler (heat) kernel $\exp\left(-t H^{\mathrm{MHO}}_\lambda\right)(x,y)$. This kernel has poles for $\lambda t \in i\pi\mathbb Z$ \cite[Prop. 2.1]{MATSUMOTO1995168}, hence we cannot hope to obtain an analytic extension beyond the open right half plane. 

To control $(h_\lambda-\mu\mathds{1})^{-1}$ localized in $U_k$, we cover $U_k$ by finitely many discs $B(x_*,\delta)$, with $\delta=\textrm{small constant}\times 2^{-k}$. On each $B(x_*,\delta)$, we rescale via the change of variables $x=x_*+\delta y$. Then, $B(x_*,\delta)$ goes over to the unit disc $B(0,1)$, and $h_\lambda-\mu\mathds{1}$ becomes a constant multiple of a second order differential operator:
\[ L=-\Delta_y+\vec{\gamma}(y)\cdot\nabla_y+\tilde{\gamma}(y), \]
whose symbol $\sigma(L)(y,\eta)= |\eta|^2+i\vec{\gamma}(y)\cdot\eta+\tilde{\gamma}(y)$ satisfies, for all $y\in B(0,1)$, $\eta\in\mathbb R^2$ and all $\alpha,\beta$,  the following estimates:
\begin{align*}
\Big|\partial_y^\alpha\partial_\eta^\beta\left[\sigma(L)(y,\eta)\right]\Big| &\le C_{\alpha\beta}\ \left[ \Lambda^2\delta^4+|\eta|^2\right]^{1-\frac{|\beta|}{2}},\\
\Big|\sigma(L)(y,\eta) \Big| &\ge c\left[ \Lambda^2\delta^4 + |\eta|^2 \right].
\end{align*}
Second order differential operators satisfying the above conditions behave like $-\Delta+\left(\Lambda^2 \delta^4\right)\mathds{1}$ and their resolvents may be controlled accordingly; in particular, their resolvents may be analytically continued. To make this precise, it is convenient to use pseudodifferential operator calculus (see, e.g. \cite{Hörmander2007}). This concludes our discussion of the case of functions supported in $U_k$. 

Combining our results on $U_{\rm inner}$ and $U_k$, we establish the desired analytic continuation of $\lambda\mapsto (h_\lambda-\mu\mathds{1})^{-1}\chi$ for ${\rm supp}(\chi)\subset B_a(0)$, $\lambda\in \Gamma_\gamma$, and $\frac12\Lambda\le|\lambda|\le2\Lambda$. Finally, combining our results for $\Lambda=2^n$ as $n$ varies, we control 
$\lambda\mapsto (h_\lambda-\mu\mathds{1})^{-1}\chi$ for $\lambda\in \Gamma_\gamma$, where  ${\rm supp}(\chi)\subset B_a(0)$. 
Our outline of the proof of analyticity of $(h_\lambda-\mu\mathds{1})^{-1}\chi$ is now complete. Our analysis of the double-well resolvent $(H_\lambda-\mu\mathds{1})^{-1}\chi$ is similar.

\subsection*{On the construction of periodic crystals with flat bands}
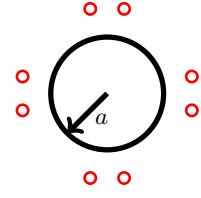
\begin{figure}
	    \centering
	    \begin{tikzpicture}[x=1cm,y=1cm,scale=0.75]
\draw [line width=2pt] (0,0) circle (1cm);
\draw [line width=1pt,red] (1.5,0.3) circle (0.1cm);
\draw [line width=1pt,red] (1.5,-0.3) circle (0.1cm);
\draw [line width=1pt,red] (-1.5,0.3) circle (0.1cm);
\draw [line width=1pt,red] (-1.5,-0.3) circle (0.1cm);
\draw [line width=1pt,red] (0.3,1.5) circle (0.1cm);
\draw [line width=1pt,red] (0.3,-1.5) circle (0.1cm);
\draw [line width=1pt,red] (-0.3,1.5) circle (0.1cm);
\draw [line width=1pt,red] (-0.3,-1.5) circle (0.1cm);


\draw [->,line width=2pt] (0,0) -- (-0.7,-0.7);
\node[below] at (-0.1,-0.2) {$a$};

\end{tikzpicture}
	    \caption{Single-well potential for flat band crystal.  }
	    \label{fig:setup for flat band single well potnetial}
	\end{figure}

Consider a single planet surrounded by eight sophons as in Figure \ref{fig:setup for flat band single well potnetial}. Now form a periodic arrangement by placing central planets at the vertices of $D^\prime \mathbb Z^2$, where $D^\prime>0$ is fixed and taken sufficiently large. By carefully adjusting the position of the sophons, we can again make the hopping coefficient vanish between any two nearest neighbors of the lattice $D^\prime \mathbb Z^2$. Thanks to [\ref{eq:crystal-limit}] this leads to a crystal structure with nearly flat band spectrum centered near $e_{\lambda}$. Further, as one varies within the family $\mathcal{F}_\lambda$, the structure can be tuned, between one that supports a flat band and one that does not. This may lead to a novel tunable platform for exploring flat band physics.

\subsection*{Open questions and conjectures}
\begin{enumerate}
\item Is $\rho_\lambda$ analytic in a half-plane $\Re\lambda\ge \lambda_0 >0$? Do the bounds of Theorem 4 hold with $\gamma =1$ for typical large real $\lambda$? 
    \item Are there double well potentials $v$, independent of $\lambda$, such that the Hamiltonian $(P+\frac12 \lambda b X^\perp)^2+\lambda^2 v(X)$ gives rise to an infinite sequence of $\lambda$'s, tending to infinity, for which the magnetic eigenvalue splitting vanishes, and another infinite sequence of $\lambda$'s for which the hopping coefficient vanishes?
%
    \item Can the transition from a symmetric to antisymmetric ground state be observed experimentally? 
    Or the flat band crystal, by the mechanism described?
    
\item Are there analogous phenomena in three dimensions? If the magnetic field is perpendicular to the displacement between the wells, then we suspect that producing counterexamples in the 3D case is actually easier than in the 2D case. However, if the magnetic field is not perpendicular to the displacement, new questions may arise.
\end{enumerate}

{\ }

     \subsection*{Acknowledgements}
 MIW was supported in part by NSF grant DMS-1908657, DMS-1937254 and Simons Foundation Math + X Investigator Award \# 376319 (MIW). Part of this research was carried out during the 2023-24 academic year, when MIW was a Visiting Member in the School of Mathematics - Institute of Advanced Study, Princeton, supported by the Charles Simonyi Endowment, and a Visiting Fellow in the Department of Mathematics at Princeton University.
The authors wish to thank Antonio Cordoba and David Huse for stimulating discussions. We thank P.A. Deift and J. Lu for their careful reading of the paper, and insightful comments and questions. 

\bibliography{magnetic-fsw.bib}

\end{document}